\documentclass[11pt]{article}
\usepackage{amsfonts}
\usepackage{epsfig}

\def\underset#1#2{\mathrel{\mathop{#2}\limits_{#1}}}

\newfont{\blackb}{msbm10 scaled\magstep1}

\def\Bbb#1{\hbox{\blackb #1}}

\newfont{\calig}{cmsy10 scaled\magstep1}

\def\text#1{\hbox{#1}}

\newtheorem{theorem}{Theorem}[section]
\newtheorem{proposition}{Proposition}[section]
\newtheorem{remark}{Remark}[section]
\newtheorem{corollary}{Corollary}[section]
\newtheorem{lemma}{Lemma}[section]

\def\be{\begin{equation}}
\def\ee{\end{equation}}
\def\ben{\begin{displaymath}}
\def\een{\end{displaymath}}
\def\baa{\begin{eqnarray}}
\def\eaa{\end{eqnarray}}

\def\ba{\begin{array}}
\def\ea{\end{array}}

\def\3{\ss}
\def\a{\alpha}
\def\b{\beta}

\def\d{\delta}

\def\l{\lambda}

\def\ph{\phi}

\def\t{\tau}
\def\th{\vartheta}
\def\Th{\Theta}

\def\ph{\phi^{\infty}}
\def\psih{\psi^{\infty}}

\def\phi{\varphi}

\def\eb{{\bf e}}

\def\Lh{\Gamma}
\def\B{{\bf B}}
\def\C{\mathbb{C}}
\def\Z{\mathbb{Z}}

\def\t0{\Theta_0}
\def\z{{\bf z}}
\def\m{{\bf m}}
\def\la{\label}
\def\Ref{\ref}
\def\c{\cite}
\def\f{\frac}
\def\L{{\cal L}}
\def\p{\partial}
\def\pb{{\bf p}}
\def\qb{{\bf q}}

\def\Ref#1{(\ref{#1})}
\def\a0{\tilde{a}}
\def\b0{\tilde{b}}
\def\c0{\tilde{c}}
\def\d0{\tilde{d}}
\def\G0{\tilde{G}}
\def\A0{\tilde{A}}
\def\tr{{\rm tr}}
\def\0{S}
\def\1{T}
\def\j{{S_j}}
\def\log{\ln}
\begin{document}

\title{On solutions of the Schlesinger Equations 
in Terms of $\Theta$-Functions}
\author{A.~V.~Kitaev 
\thanks{$\;\;$E-mail: kitaev@pdmi.ras.ru .
Supported by Alexander von Humboldt Foundation}\hskip0.5cm and\hskip0.5cm 
D.~A.~Korotkin
\thanks{$^*$E-mail: korotkin@aei-potsdam.mpg.de}
\addtocounter{footnote}{-1}
\\
\\Max-Planck-Institut f\"ur Gravitationsphysik, Albert-Einstein-Institut,\\
Schlaatzweg 1, D-14473 Potsdam, Germany\\
and\\
Steklov Mathematical Institute, Fontanka, 27, St.Petersburg 191011 Russia}

\maketitle
\begin{abstract}
In this paper we construct explicit solutions and calculate the corresponding
$\tau$-function to the system of Schlesinger equations describing isomonodromy deformations of $2\times 2$ matrix linear ordinary differential equation whose
coefficients are rational functions with poles of the first order; in particular, in the case when the coefficients have four poles of the first order and the corresponding Schlesinger system reduces to the sixth Painlev\'e equation with the parameters $1
/8,\,-1/8,\,1/8,\,3/8$, our construction leads to a new representation of the general solution to this Painlev\'e equation obtained earlier by K.~Okamoto and N.~Hitchin, in terms of elliptic theta-functions.\vspace{24pt}


\vspace{24pt}\\
Short title: On $\Theta$-Function Solutions of Schlesinger Equations 
\end{abstract}

\newpage
\section{Introduction}
\setcounter{equation}{0}
The Schlesinger equations \cite{S} arise in the context of the 
following Riemann-Hilbert (inverse monodromy)
problem:\\
for an arbitrary $g\in\Bbb N$ and distinct $2g+2$ points
$\lambda_j\in\Bbb C$, construct a function $\Psi(\lambda):
\Bbb{CP}^1\setminus\{\lambda_1,\dots,\lambda_{2g+2}\}\to {\rm SL}(2,\Bbb C)$
which has the following properties;\\
1) $\Psi(\infty)=I$,\\
2) $\Psi(\lambda)$ is holomorphic for all
$\lambda\in\Bbb{CP}^1\setminus\{\lambda_1,\dots,\lambda_{2g+2}\}$,\\
3) $\Psi(\lambda)$ has regular
singular points  at $\lambda=\lambda_j,\;j=1,\dots,2g+2$, with given
monodromy matrices, $M_j\in{\rm SL}(2,\C)$ \\
In the case when the monodromy matrices are independent of the parameters $\lambda_1,\dots,\lambda_{2g+2}$, the function $\Psi\equiv\Psi(\lambda)$ 
solves the following matrix differential equation,
\be
\frac{d\Psi}{d\lambda}=\sum\limits_{j=1}^{2g+2}
\frac{A_j}{\lambda-\lambda_j}\Psi,
\la{PsiAA}\ee
where the $sl(2,\C)$-valued matrices $A_j$ solve 
the system of Schlesinger equations,
\be
\frac{\partial A_j}{\partial\lambda_i}=
\frac{[A_i,A_j]}{\lambda_i-\lambda_j},\;\;\;i\neq j\;,\;\;\;\;\;
\frac{\partial A_i}{\partial\lambda_i}=
-\sum_{j\neq i}
\frac{[A_i,A_j]}{\lambda_i-\lambda_j}.
\label{sch} 
\ee
Obviously, the eigenvalues of $A_j$,which will be denoted by $\f{t_j}{2}$ and
$-\f{t_j}{2}$ in the sequel, are integrals of motion of system (\ref{sch}).

The important object associated with system (\ref{sch}) is the so-called
$\tau$-function - the function generating  Hamiltonians of the
Schlesinger  system \cite{SMJ,JMU,JM}; it can by defined as the solution to the following system of equations, 
\ben
\f{\partial\log\tau}{\partial\l_j}\equiv\sum_{i\neq j}\frac{{\rm tr}
A_jA_i}{\lambda_j-\lambda_i}
\een
(see Sec.2 for details).
 
For $g=1$, the Schlesinger system may equivalently be rewritten
in terms of a single function of one variable, the position $y(t)$ of the zero 
of the $(12)$ matrix element
of the function $\f{A_1}{\l}+\f{A_2}{\l-1}+\f{A_3}{\l-t}$
in the $\l$-plane. The equation for $y(t)$ turns out to coincide with 
the sixth Painlev\'e equation,
\begin{eqnarray}
  \label{eq:P6}
  &\frac{d^2y}{dt^2}=\frac 12\left(\frac 1y+\frac 1{y-1}+\frac 1{y-t}\right)
\left(\frac{dy}{dt}\right)^2-\left(\frac 1t+\frac 1{t-1}+\frac 1{y-t}\right)
\frac{dy}{dt}+&\nonumber\\
&\frac{y(y-1)(y-t)}{t^2(t-1)^2}\left(\alpha+\beta\frac t{y^2}+
\gamma\frac{t-1}{(y-1)^2}+\delta\frac{t(t-1)}{(y-t)^2}\right),&
\end{eqnarray}
where
\begin{equation}
  \label{DEF_T_J}
\alpha\equiv\frac{(t_1-1)^2}2,\quad\beta\equiv-\frac{t_2^2}2,\quad
\gamma\equiv\frac{t_3^2}2,\quad\delta\equiv\frac 12-\frac {t_4^2}2,
\end{equation}

K.~Okamoto showed \cite{O} that
the general solution to the sixth Painlev\'e equation  can be written 
explicitly in terms of elliptic functions provided that the set of
the parameters $t_j$ satisfy one of the following conditions:
$t_i\in\Bbb Z,\;\;t_1+\dots+t_4\in 2\Bbb Z$ or $t_i+\f12\in \Z$.
More recently, the algebro-geometric aspects of the sixth Painlev\'e 
equation have once again attracted the attention, see the papers \cite{H,M} 
(some details which are relevant to our work are given in Appendix).

Our interest to the problem  of finding explicit solutions of the Schlesinger 
system in algebro geometric terms 
was initiated, on one hand, by the work of Okamoto,
and, on the other hand, by our papers \cite{Kor,KorMat,Kor1,K}, 
devoted to the study of 
solutions to the Ernst equation arising as a partial case 
of the vacuum Einstein equations; in particular, it turns out that
some of the elliptic solutions of the Ernst equation studied in \cite{Kor1} 
may also be described by the sixth Painlev\'e equation \cite{K}: 
in fact, being rewritten 
in appropriate variables, these solutions give a
certain one-parameter sub-family of the Okamoto's solutions with $t_j=1/2$.

In this paper we solve, in terms of
theta-functions, the
inverse monodromy problem  formulated at the beginning of the Introduction
for an arbitrary $g$ and an arbitrary
set of  anti-diagonal monodromy matrices. 
Our approach  originated from the so-called finite-gap integration method
for the integrable systems \cite{DMN}.
The solution of the inverse monodromy problem allows, in turn,
to express in terms
of theta functions the
$2g$-parameter family of solutions to the Schlesinger system for $t_j=\f12$ 
and  calculate  
the corresponding $\mathbf{\tau}$-function.
In contrast to the common belief 
(which got its origin in the papers \cite{JM,SMJ,JMU}) 
that for algebro geometrical solutions to integrable systems
the $\tau$-function simply coincides with certain theta functions,
in the present case,
the $\tau$-function (up to multiplication by an arbitrary constant) is given by
the following expression,
\begin{equation}
  \label{TAU_MAIN}
\tau(\{\l_j\})=\f{\Theta[{\bf p},{\bf q}](0|\B)}{\sqrt{\det{\cal A}}}
\prod\limits_{j<k}(\lambda_j-\lambda_k)^{-\frac 18},
\end{equation}
where the vectors ${\bf p}\in \C^g,\,{\bf q}\in\C^g$
are parameters corresponding to parameters of the monodromy matrices, $\B$
is the matrix of $b$-periods of the hyperelliptic curve
\ben
w^2=\prod_{j=1}^{2g+2}(\l-\l_j),
\een
and
\ben
{\cal A}_{kj}\equiv 2\int_{\l_{2j+1}}^{\l_{2j+2}}\f{\l^{k-1} d\l}{w}\;,\;\;\;
j,k=1,\dots,g.
\een
For the elliptic case $g=1$, 
applying a conformal transformation of the $\l$-plane, one can always map the points $\l_1,\dots,\l_4$ to $0,1,t$ and $\infty$, respectively ($t$ is equal to
the cross-ratio of the points $\l_1,\dots,\l_4$). Then (again up to an arbitrary constant) the
 $\mathbf{\tau}$-function (\ref{TAU_MAIN}) can be
rewritten in the following form,
\begin{equation}
  \label{TAU_PAINLEVE_6}
\mathbf{\tau}(t)=\f{\theta_{p,q}(0\vert\sigma)}{\sqrt[8]{t(t-1)}}
\left[\int_0^1\f{d\l}{\sqrt{\l(\l-1)(\l-t)}}\right]^{-\f12},
\end{equation}
where $\theta_{p,q}(0\vert\sigma)$ is the elliptic theta-function with 
characteristic $[p,q]$: here, the module $\sigma(t)$ of the curve
$w^2=\l(\l-1)(\l-t)$ is chosen so that 
$t=\theta_4^4(0|\sigma)/\theta_2^4(0|\sigma)$.

The latter $\mathbf{\tau}$-function defines a new representation
of the  solution
to the sixth Painlev\'e equation with the parameters $t_j=1/2$ i.e.
\begin{equation}
\label{eq:OKAMOTO_COEFFICIENTS}
\alpha=\frac 18,\quad\beta=-\frac 18,\quad
\gamma=\frac 18,\quad\delta=\frac 38\;:  
\end{equation}

\begin{equation}
  \label{P6_SOLUTION_IN_TAU}
  y(t)=t-t(t-1)\left[D\left(\frac{\frac d{dt}D(\mathbf{\tau})}
  {\frac d{dt}D\left(\sqrt[8]{t(t-1)}\mathbf{\tau}\right)}\right)
+\frac{t(t-1)}{D^2\left(\sqrt[8]{t(t-1)}\mathbf{\tau}\right)}\right]^{-1},
\end{equation}
where the operator $D$ is defined as follows,
$$
D(\cdot)\equiv t(t-1)\frac d{dt}\log(\cdot).
$$

As a corollary of sixth Painlev\'e equation (\ref{eq:P6}) with coefficients 
(\ref{eq:OKAMOTO_COEFFICIENTS}), function 
\ben
\zeta(t)\equiv D(\tau)
\een
where the $\tau$-function $\tau(t)$ is given by (\ref{TAU_PAINLEVE_6}),
satisfies the following equation:
\be
[t(t-1)\zeta'']^2=\zeta'[(\zeta'+\f14)^2-((2t-1)\zeta'-\zeta)^2]
\la{zeta}\ee

One more form of the solution (\ref{P6_SOLUTION_IN_TAU}), namely, 
\begin{equation}
  \label{eq:ANDERE_FORM}
  y(t)=\frac{tu(\frac\sigma 2|\sigma)}{u(\frac\sigma 2,|\sigma)+
(1-t)u(\frac 12|\sigma)},\quad {\rm where}\quad u(z|\sigma)=\frac
{\frac\partial{\partial z}\log\frac\partial{\partial z}\log
\frac{\theta_{p,q}(z|\sigma)}{\theta_1(z|\sigma)}}
{\frac\partial{\partial\sigma}\log\frac{\theta_{p,q}(z|\sigma)}
{\theta_1(z|\sigma)}},
\end{equation}
may be obtained
from our construction by a straightforward calculation of the position of the
zero of the $(12)$ component of the matrix $\Psi_\l\Psi^{-1}$ in the $\l$-plane.

This paper is organized as follows.
In Section 2, we recall some basic facts about isomonodromy deformations
and Schlesinger equations.
In Section 3, we begin with the
solution of an inverse monodromy problem with an arbitrary even number 
of singular points and anti-diagonal monodromy matrices.
In Section 4, we find the related $\tau$-function, and finally, in 
Section 5, we apply the results of the previous sections to the $g=1$ case,
i.e., to the sixth Painlev\'e equation. 

It is also worth mentioning, that the solution of some
inverse monodromy 
problems, including singularities of regular and irregular type in the
framework of the finite-gap integration technique, were given by M.~Jimbo
and T.~Miwa \cite{JM}, however, their construction can not be applied to solve
the inverse monodromy problems considered here. In the
case of $2\times 2$ monodromy problems with only regular singularities, say,
the construction by Jimbo and Miwa leads to an
analytic function with $3g+2$ regular singular points whose $2g+2$ monodromy
matrices, after a proper normalization (see Section \ref{SECTION2}), equal 
$i\sigma_1$, and $g$ monodromy matrices are just equal to $-I$. 
Therefore, the solution of the Schlesinger system, which can be obtained from the construction of Jimbo and Miwa, does not contain any
parameters in contrast to the construction presented in this paper.

Simultaneously with the present work, solution of the same Riemann-Hilbert problem was given in the paper
of P.Deift, A.Its, A.Kapaev and X.Zhou \cite{DIKZ} in rather different terms. The problem of
calculation of corresponding $\tau$-function (\ref{TAU_MAIN}) was not considered there.

\section{The Schlesinger Equations}\label{SECTION2}
\setcounter{equation}{0}
In this section we recall the basic notations and definitions related to 
isomonodromy deformations of $2\times 2$ matrix linear ordinary 
differential equation,
\begin{equation}\label{AEQ}
\frac d{d\lambda}\Psi=A(\lambda)\Psi,
\end{equation} 
where $A(\lambda)\in sl_2(\Bbb C)$ is a rational function of $\lambda$
with $2g+2$ poles of the first order,
\begin{equation}\label{AKDEF}
A(\lambda)=\sum\limits_{j=1}^{2g+2}\frac{A_j}{\lambda-\lambda_j},\;\;\;
i\neq j\Rightarrow \lambda_i\neq\lambda_j,\;\;\;\frac d{d\lambda}A_j=0.
\end{equation}
We suppose that $\lambda=\infty$ is not a pole, which means that
the following condition is fulfilled
\begin{equation}
  \label{eq:INFTY_REGULAR}
  \sum\limits_{j=1}^{2g+2}A_j=0.
\end{equation}
To fix a fundamental solution of Eq.~(\ref{AEQ}), choose a point
$\lambda_0\in\Bbb P\setminus\{\lambda_1,\dots,\lambda_{2g+2}\}$ and impose
the following normalization condition:
\begin{equation}
  \label{NORM1}
\Psi(\lambda_0)=I.
\end{equation}
Since $\tr A(\l)=0$, this means that $\det\Psi(\lambda)=1$ 
for $\lambda\in\Bbb C$.
Now one defines the {\it monodromy matrices},
$$
M_j=\Psi(\lambda_0)\left\vert_{\gamma_j}\right.,\;\;\;
k=j,\dots, 2g+2,
$$ 
as analytic continuations of
the fundamental solution normalized by condition (\ref{NORM1}) along
the generators, $\gamma_k$, of the fundamental group
$\pi_1(\Bbb{CP}^1\setminus\{\lambda_1,\lambda_2,\dots,
\lambda_{2g+2}\},\,\lambda_0)$ defined in the figure 1.


The monodromy matrices satisfy the cyclic relation,
\begin{equation}
  \label{eq:CYCLIC}
M_{2g+2}\cdot\dots\cdot M_1=I,
\end{equation}
and generate a subgroup of $SL(2,\Bbb C)$, i.e.,
\begin{equation}
  \label{eq:DET_M_K}
  \det M_j=1,\;\;\;j=1,\dots,2g+2.
\end{equation}
Matrix elements of $M_j$ and eigenvalues $\pm\frac{t_j}2$ of the matrices
$A_j,\;j=1,\dots,2g+2$, are called the {\it monodromy data} of the function 
$\Psi$. 
The monodromy data are locally analytic functions of the variables 
$A_1,\dots,A_{2g+2}$ and
$\lambda_0,\lambda_1,\dots,\lambda_{2g+2}$. The condition
\begin{equation}
  \label{eq:ISO_CONDITION}
  \frac{dt_j}{d\lambda_l}=0\;\;\;{\rm and}\;\;\;\frac{dM_j}{d\lambda_l}=0,
\;\;\;{\rm for}\;j,\,l=1,\dots,2g+2,
\end{equation}
is called the {\it isomonodromy condition}. The isomonodromy condition 
(\ref{eq:ISO_CONDITION}) is equivalent
to the following system of linear differential equations for the function $\Psi$:
\begin{equation}
  \label{eq:PSI_ISO}
  \frac{d\Psi}{d\lambda_j}=\left(\frac{A_j}{\lambda_0-\lambda_j}
-\frac{A_j}{\lambda-\lambda_j}\right)\Psi,\;\;\;\;\;\;j=1,\dots,2g+2.
\end{equation}
Following \cite{S} we choose the normalization point $\lambda_0=\infty$
to exclude the nonessential parameter $\lambda_0$. In this case the 
compatibility condition of system (\ref{eq:PSI_ISO}), (\ref{AEQ})
reads as the following system of nonlinear ODEs, the 
{\it Schlesinger equations}:
\begin{eqnarray}\label{SCHLESINGER}
j\neq i:& & \frac{\partial A_j}{\partial\lambda_i}=
\frac{[A_i,A_j]}{\lambda_i-\lambda_j},
\label{FIRST_SCHLESINGER}\\
j=i:& &\frac{\partial A_i}{\partial\lambda_i}=
-\sum\limits_{\underset{j\neq i}{j=1}}^{2g+2}
\frac{[A_i,A_j]}{\lambda_i-\lambda_j},
\label{LAST_SCHLESINGER} 
\end{eqnarray}
Solutions of these equations 
define {\it isomonodromy deformations} of the matrix 
elements of $A_j$. Note that system (\ref{FIRST_SCHLESINGER}),
(\ref{LAST_SCHLESINGER}) is equivalent to system (\ref{FIRST_SCHLESINGER}),
(\ref{eq:INFTY_REGULAR}).
\begin{proposition}\label{PROP_SCHLESINGER}
If a set $\{A_1,\dots,A_{2g+2}\}$ is a solution of the system 
{\rm (\ref{FIRST_SCHLESINGER}), (\ref{LAST_SCHLESINGER})}, then 
the monodromy data of the function $\Psi$, which solves 
Eq.~{\rm (\ref{AEQ})} with the corresponding matrix $A(\lambda)$ given by 
Eq.~{\rm (\ref{AKDEF})}, are independent of $\lambda_1,\dots,\lambda_{2g+2}$. 
\end{proposition}
The set of the monodromy data, $\{t_1,\dots,t_{2g+2},
\,M_1,\dots,M_{2g+2}\}\in\Bbb C^{2g+2}\times{\cal M}_{2g+2}$, where
the variety 
${\cal M}_{2g+2}\equiv{\cal M}_{2g+2}(t_1,\dots,t_{2g+2})$ is defined 
via Eqs.~(\ref{eq:CYCLIC}) and (\ref{eq:DET_M_K}), 
is known to be in one-to-one correspondence with the solutions of the
system of Schlesinger equations (\ref{FIRST_SCHLESINGER}), 
(\ref{LAST_SCHLESINGER}). The nontrivial part of this statement follows 
from the solvability of the inverse monodromy problem (see \cite{B}).
 
In this paper we consider the case when all 
$t_j=1/2$, so that the matrices $A_j$ and $M_j$ 
can be represented in the following form,
\begin{equation}
  \label{eq:A_K_IN_G_K}
  A_j=\frac 14G_j\sigma_3 G_j^{-1},\;\;\;\;M_j=iC_j^{-1}\sigma_3 C_j,
\end{equation}
and $\l$-independent matrices $ G_j$ and $C_j$ are defined via the 
asymptotic behavior of the function $\Psi$ in the neighborhood of 
the points $\lambda_j$,
\begin{equation}
  \label{eq:PSI_IN_LAMBDA_K}
\Psi\underset{\lambda\to\lambda_j}= 
(G_j+{\cal O}(\lambda-\lambda_j))(\lambda-\lambda_j)^{\frac 14\sigma_3}C_j;
\end{equation}
$$
\det G_j=\det C_j=1.
$$
In the isomonodromy case one can always choose $C_j$ to be independent of
$\lambda_1,\dots,\lambda_{2g+2}$. Hereafter we use the standard notation 
for the Pauli matrices:
$$
\sigma_1=
\left(\begin{array}{cc}
0&1\\
1&0
\end{array}\right),\;\;\;\;
\sigma_2=
\left(\begin{array}{cc}
0&-i\\
i&0
\end{array}\right),\;\;\;\;
\sigma_3=
\left(\begin{array}{cc}
1&0\\
0&-1
\end{array}\right).
$$

One can formulate the following 
\begin{proposition}\label{GENERAL}
Let $\Psi^*(Q)$ be a holomorphic function on the universal covering,
${\rm pr:}\;X\rightarrow 
\Bbb{CP}^1\setminus\{\lambda_1,\dots,\lambda_{2g+2}\},$
which has the asymptotic behavior as $\lambda={\rm pr}\,Q\to\lambda_j$ 
prescribed by
Eq.~{\rm (\ref{eq:PSI_IN_LAMBDA_K})} and normalized as $\Psi^*(Q_0)=I$ at some
point $Q_0,\;{\rm pr}\,Q_0=\lambda_0$. Then the function 
$\Psi(\lambda)=\Psi^*(Q)\vert_{{\rm pr}\,Q=\lambda}$ has the monodromy data
corresponding to the variety 
${\cal M}_{2g+2}(\pm\frac 12,\dots,\pm\frac 12)$, with the matrices $M_j$
defined via the second equation {\rm (\ref{eq:A_K_IN_G_K})}, and solves the
system  of differential equations {\rm (\ref{AEQ}), (\ref{eq:PSI_ISO})}, 
where the matrix $A(\lambda)$ is defined by Eqs.~{\rm (\ref{AEQ})} and 
{\rm (\ref{AKDEF})}.
\end{proposition}

If a set of matrices $\{A_1,\dots,A_{2g+2}\}$ is a solution of 
the system (\ref{FIRST_SCHLESINGER}), (\ref{LAST_SCHLESINGER}), then 
for any matrix $K\in{\rm SL}(2,\Bbb C)$ independent of $\lambda_1,\dots,
\lambda_{2g+2}$ the new set $\{A_j^{new}\!=KA_jK^{-1},\;\; j=1,\dots,2g+2\}$ 
is also a solution of the system. This gauge transformation on the
set of the solutions of the Schlesinger system corresponds to the
following gauge transformation of the function $\Psi(\l)$, 
\begin{equation}
  \label{GAUGE}
\Psi^{new}=K\Psi K^{-1},
\end{equation} 
which leaves the normalization condition (\ref{NORM1})  invariant and
acts on ${\cal M}_{2g+2}$ in the same way as on the space of the 
solutions, 
\begin{equation}\label{CONJUGATION}
M_j^{new}=KM_jK^{-1}.
\end{equation} 
By choosing $K=C_0C_1$, where
$C_1$ is given by (\ref{eq:PSI_IN_LAMBDA_K}) for $j=1$ and
$C_0=\frac i{\sqrt 2}(\sigma_3+\sigma_1)$, 
we use this gauge transformation to fix 
\begin{equation}
  \label{eq:NORM2}
M_1=i\sigma_1.
\end{equation}
Since we have one more parameter in our gauge transform, 
$C_0\to C_0\kappa^{\sigma_3}$, we can use the remaining freedom to remove
one more parameter from ${\cal M}_{2g+2}$. More exactly, by making one
more gauge transform (\ref{GAUGE}) with the matrix 
$K=C_0\kappa^{\sigma_3}C_0^{-1}$, we, by choosing appropriately the 
parameter $\kappa$, fix the next monodromy matrix $M_2$:\\
If ${\rm tr}\,(M_2\sigma_1)^2\neq -2$, then 
\begin{equation}
  \label{eq:NORM3}
M_2=\left(\begin{array}{cc}
0&m_2\\
-m_2^{-1}&0\end{array}\right),\quad 
m_2\in\Bbb C^*\equiv\Bbb C\setminus\{0,\infty\};
\end{equation}
if ${\rm tr}\,(M_2\sigma_1)^2=-2$ but $M_2\neq\pm i\sigma_1$, then
$M_2=\pm i(\sigma_3+\sigma_1+i\sigma_2)$; and, finally, if 
$M_2=\pm i\sigma_1$, then we can use the parameter $\kappa$ to fix 
analogously the structure of the next matrix, $M_3$.

The variety ${\cal M}_{2g+2}(\pm\frac 12,\dots,\pm\frac 12)$ contains 
the following sub-variety, 
${\Bbb C}_{2g}^*\cong\Bbb T_{2g}\times\Bbb R^{2g}$:
\begin{equation}
  \label{eq:CYLINDERS}
M_j=\left(\begin{array}{cc}
0&m_j\\
-m_j^{-1}&0\end{array}\right),\quad j=1,\dots,2g+2,
\end{equation}
where 
\begin{equation}
  \label{eq:CYLINDERS_CONDITIONS}
  m_1=i,\quad m_j\in\Bbb C^*,\;\;\;j=2,\dots,2g+2;\quad\quad
\prod\limits_{j=1}^{g+1}m_{2j}=(-1)^{g+1}\prod\limits_{j=1}^{g+1}m_{2j-1}.
\end{equation}
Note that if the matrices $M_1$ and $M_2$ are fixed by Eqs.~(\ref{eq:NORM2})
and (\ref{eq:NORM3}) correspondingly, then
$\dim_{\Bbb C}{\cal M}_{2g+2}(\pm\frac 12,\dots,\pm\frac 12)=4g-2$ and
$\dim_{\Bbb C}\C_{2g}^*=2g$;
in fact, for $g=1$ the sub-variety $\C_{2g}^*$ constitutes almost all 
the variety ${\cal M}_{2g+2}(\pm\frac 12,\dots,\pm\frac 12)$. More precisely,
one can formulate the following
\begin{proposition}{\rm\cite{H}}\label{HITCHIN_M_4}
If $g=1$, then the variety 
${\cal M}_{4}(\pm\frac 12,\dots,\pm\frac 12)$ coincides, up to the
conjugation defined by Eq.~{\rm (\ref{CONJUGATION})} with arbitrary matrix 
$K\in{\rm SL}(2,\Bbb C)$, with the union of the following two sets of the
monodromy matrices:
\begin{equation}
  \label{eq:SET_GENERAL}
1)\quad M_k=\left(\begin{array}{cc}
0&m_k\\
-\frac 1{m_k}&0
\end{array}\right)\!,\;\,k=1,\dots\,4,\;\;\;m_1=i,\;\;
m_k\in\Bbb C^*,\;\;\;m_4m_2=im_3;
\end{equation}
\begin{equation}
  \label{eq:SET_SPECIAL}
2)\quad  M_1=-i\sigma_3,\;\,
M_2=i\epsilon_2\!\left(\!\!\!\begin{array}{cc}
-1&a-1\\
0&1
\end{array}\!\!\!\right)\!\!,\;\;
M_3=i\epsilon_3\!\left(\!\!\!\begin{array}{cc}
-1&a\\
0&1
\end{array}\!\!\!\right)\!\!,\;\,
M_4=i\epsilon_4\!\left(\!\!\!\begin{array}{cc}
-1&1\\
0&1
\end{array}\!\!\!\right)\!\!,\!\!\!
\end{equation}
where
$
\epsilon_k=\pm 1,\;\;\epsilon_2\epsilon_3\epsilon_4=1,\;\;a\in\Bbb C.
$
\end{proposition}
Isomonodromy deformations of Eq.~(\ref{AEQ}) in the case when the matrix
$A(\lambda)$ has four poles are governed by solutions to the sixth 
Painlev\'e equation (\ref{eq:P6}). 
Here we rewrite the corresponding relation 
given by M.~Jimbo and T.~Miwa \cite{JM} in the notation which more suits to 
our basic construction:\\
Denote by $\vec g_j^{\,p}$ the $p$th column of the matrix $G_j$ from 
Eq.(\ref{eq:A_K_IN_G_K}), 
and introduce new matrices
$G_{ij}^{pq}\stackrel{{\rm def}}{=}(\vec g_i^{\,p}\,\vec g_j^{\,q})$;
in particular, $G_{jj}^{12}\equiv G_j$.
\begin{proposition}\label{SCH_P6}
The functions
\begin{equation}
  \label{eq:B_J}
\hat A_j^{12}=t_j
\frac{\det G_{j1}^{12}\det G_{1j}^{22}}{\det G_{11}^{12}\det G_{jj}^{12}},\;\;
\;\;j=1,\dots,4,
\end{equation}
depend on the variables $\{\l_k\}$ only through their cross-ratio, 
\begin{equation}
  \label{eq:DEF_T}
t=\frac{\lambda_3-\lambda_1}{\lambda_3-\lambda_2}\;
\frac{\lambda_4-\lambda_2}{\lambda_4-\lambda_1}.  
\end{equation}
Moreover, the function
\begin{equation}
  \label{eq:DEF_SOLUTION}
  y(t)=-\frac t{1+(1-t)\hat A_4^{12}/\hat A_2^{12}}=
\frac 1{1-\frac{1-t}t\hat A_3^{12}/\hat A_2^{12}}
\end{equation}
is the solution of the sixth Painlev\'e equation (\ref{eq:P6}) with 
the parameters given by Eq.(\ref{eq:OKAMOTO_COEFFICIENTS}).

\end{proposition}
{\it Proof}. If the set $\{A_j\}$ is a solution of the system (\ref{FIRST_SCHLESINGER}), (\ref{eq:INFTY_REGULAR}), then the
monodromy data of the function $\Psi$, which solves the corresponding Eq.~(\ref{AEQ}), are independent of 
$\{\l_j\}$ and $\l$. Define the new variable
\begin{equation}
  \label{MU}
\mu=\frac{\lambda_3-\lambda_1}{\lambda_3-\lambda_2}\;
\frac{\lambda-\lambda_2}{\lambda-\lambda_1}
\end{equation}
and consider
\begin{equation}
  \label{eq:PHI_DEF}
\hat\Psi=G_1^{-1}\Psi C_1^{-1}
\end{equation}
as a function of $\mu$. In the complex $\mu$-plane the function $\Phi$
has singularities only at the points $0,\,1,\,t$, and $\infty$ 
with the behavior prescribed by Eqs.~(\ref{eq:PHI_DEF}) and (\ref{eq:PSI_IN_LAMBDA_K}): it is normalized at $\mu=\infty$ by the
condition
$$
\hat\Psi\underset{\mu\to\infty}=
\left(I+{\cal O}\left(\mu^{-1}\right)\right)\mu^{\f14\sigma_3},
$$
and its monodromy data are independent of  
$\{\l_j\}$. Such a function is uniquely defined and depends on 
$\{\l_j\}$ only via the cross-ratio $t$.
It means that the logarithmic derivative,
\begin{equation}
  \label{HAT_A_MU_DEF}
\frac{d\hat\Psi}{d\mu}\hat\Psi^{-1}=\frac{\hat A_2}\mu+\frac{\hat A_3}{\mu-1}+
\frac{\hat A_4}{\mu-t}\stackrel{{\rm def}}=\hat A(\mu),
\end{equation}
and, in particular, the matrices
$$
\hat A_j=\frac{t_j}2G_1^{-1}G_j\sigma_3G_j^{-1}G_1
$$
also depend on $\{\l_j\}$ only via the variable $t$. The matrices 
$\hat A_j$ can be rewritten as follows,
$$
\hat A_j=-\frac{t_j}4\hat G_j^{-1}\sigma_3\hat G_j, 
$$
where
\begin{equation}
  \label{eq:HAT_G_J_DEF}
  \hat G_j=\left(\begin{array}{cc}
\det G_{j1}^{11}&\det G_{j1}^{12}\\
\det G_{j1}^{21}&\det G_{j1}^{22}
\end{array}\right),\quad\quad
\det\hat G_j=\det G_j\,\det G_1.
\end{equation}
To complete the proof one has to recall that according to \cite{JM}
the function $y(t)$, which solves the equation
$\hat A^{12}(y)=0$, where $ A^{12}(\cdot)$ is the corresponding matrix
element of $\hat A(\cdot)$ (see (\ref{HAT_A_MU_DEF})), is the solution 
of the sixth Painlev\'e equation.
\vskip0.5cm
\begin{remark}\rm
Proposition \ref{SCH_P6} is valid not only for the present case, when all coefficients $t_j$ equal to $\frac 12$,
 but also in the case of arbitrary complex $t_j$. In the latter case the
function $y(t)$ (\ref{eq:P6}) solves the sixth Painlev\'e equation with the
coefficients:
$$
\alpha=\frac 12(t_1-1)^2,\;\;\;\beta=-\frac 12t_2^2,\;\;\;
\gamma=\frac 12t_3^2,\;\;\;\delta=\frac 12(1-t_4^2).
$$ 
\end{remark}
\vskip0.5cm

The object playing the important role in  applications
of isomonodromy deformations in differential geometry and mathematical
physics is the so-called tau function $\tau(\{\l_j\})$. We recall 
here the definition of the $\bf\tau$-function given in
\cite{JM,SMJ,JMU}.\\
The Schlesinger equations (\ref{FIRST_SCHLESINGER}), 
(\ref{LAST_SCHLESINGER}) can be rewritten in the Hamiltonian form,
\begin{equation}
  \label{SCHLESINGER_HAMILTONIAN}
\frac{dA_j}{d\lambda_k}=\{H_k,\,A_j\},
\end{equation}
where the Poisson bracket is defined as follows,
\begin{equation}
  \label{POISSON_STRUCTURE}
\{(A_i)_{ab},\,(A_j)_{cd}\}=\delta_{ij}\left((A_i)_{ad}\delta_{cd}-
(A_i)_{bc}\delta_{ad}\right), 
\end{equation}
and the Hamiltonians are given by
\begin{equation}
  \label{DEF_H_J}
H_j=\frac 12\underset{\lambda=\lambda_j}
{\rm Res}{\rm Tr}A^2(\lambda)=-\underset{\lambda=\lambda_j}
{\rm Res}\det A(\lambda)\equiv\sum\limits_{i\neq j}^{2g+2}\frac{{\rm tr}
A_jA_i}{\lambda_j-\lambda_i}.
\end{equation} 
One proves that
\begin{equation}
  \label{H_RELATIONS}
\{H_k,\,H_j\}=0,\;\;\;\;\quad\frac{\partial H_k}{\partial\lambda_j}=
\frac{\partial H_j}{\partial\lambda_k},
\end{equation}
which imply the compatibility of system
(\ref{SCHLESINGER_HAMILTONIAN}). Taking into account the 
previous equations one can correctly define the $\tau$-function ${\bf\tau}\equiv{\bf\tau}
(\lambda_1,\dots,\lambda_{2g+2})$ generating Hamiltonians $H_j$ by 
\be
\frac d{d\lambda_j}\log{\bf\tau}=H_j,
\la{deftau}\ee                       
which is holomorphic outside of the hyperplanes $\lambda_j=\lambda_i,\;
i,\,j=1,\dots,2g+2$.

\section{Solutions of the Schlesinger System}
\setcounter{equation}{0}

Consider the hyperelliptic curve $\L$ of genus $g$ defined by the equation
\be
w^2=\prod_{j=1}^{2g+2}(\l-\l_j)
\la{L}\ee
with arbitrary non-coinciding 
$\l_j\in\C$ and the basic cycles $(a_j,b_j)$ chosen according to 
figure~ 2.


Let us denote the fundamental polygon of $\L$ by $\hat{\L}$.
The basic  holomorphic 1-forms on $\L$ 
are given by
\be
dU^0_k=\f{\l^{k-1}d\lambda}{w},\;\;\;\;\;\;k=1,\dots,g.
\la{dUk0}\ee
Let us define $g\times g$ matrices of $a$- and $b$-periods 
of these 1-forms by
\be
{\cal A}_{kj}=\oint_{a_j}dU^0_k,\;\;\;\;\;\;\;
{\cal B}_{kj}=\oint_{b_j}dU^0_k.
\la{AB}\ee
Then the holomorphic 1-forms 
\be
dU_k=\f{1}{w}\sum_{j=1}^g ({\cal A}^{-1})_{kj} \l^{j-1} d \lambda
\la{dUk}\ee
satisfy the normalization conditions
$\oint_{a_j} dU_k=\delta_{jk}$. 

The matrices ${\cal A}$ and ${\cal B}$ define the symmetric 
$g\times g$ matrix of $b$-periods of the curve $\L$:
$$\B= {\cal A}^{-1}{\cal B}\;.$$ 
Let us now introduce the theta function with characteristic $[\pb,\qb]$ ($\pb\in\C^g$, $\qb\in\C^g$) by the following series, 
\be
\Th[\pb,\qb](\z|\B)=\sum_{\m\in \Z^g} \exp\{\pi i \langle\B(\m+\pb),\m+\pb\rangle +2\pi i \langle\z+\qb,\m+\pb\rangle\},
\la{theta}\ee
for any $\z\in\C^g$. It possesses 
the following periodicity properties:
\be
\Th[\pb,\qb](\z+\eb_j)=
e^{2\pi i p_j} \Th[\pb,\qb](\z),
\la{pera}\ee
\be
\Th[\pb,\qb](\z+\B \eb_j)=e^{-2\pi i q_j}e^{-\pi i \B_{jj}-2\pi i \z_j}
\Th[\pb,\qb](\z),
\la{perb}\ee
where
\be
\eb_j\equiv (0,\dots,1,\dots,0)
\ee
($1$ stands in the $j$th place). 

Denote the universal covering of $\L$ by  $\Lh$. 
The multi-valued on $\L$, and single-valued on $\Lh$, map 
$U(P)\in \C^g$ is defined by the contour integral
$U_j(P)=\int_{\l_1}^P dU_j$.
The vector of Riemann constants corresponding to our choice of the initial point of the map reads as follows \cite{Fay}:
\be
K=\f{1}{2}\B(\eb_1+\dots+\eb_g) +\f{1}{2}(\eb_1+2\eb_2\dots+g\eb_g).       
\la{K}\ee

The characteristic with components $\pb\in\C^g/2\C^g$,
$\qb\in\C^g/2\C^g$ is called half-integer characteristic: the half-integer 
 characteristics are in one-to-one correspondence with the half-periods
$\B\pb+\qb$.
If the scalar product 
$4\langle\pb,\qb\rangle$ is odd, then the related theta function is 
odd with respect to its argument $\z$ and the characteristic $[\pb,\qb]$ is called odd, and if this scalar product is even, then the theta function $\Th[\pb,\qb](\z)$ is even with respect to $\z$ and the characteristic $[\pb,\qb]$ is called even.

The odd characteristics which will be of importance for us in the sequel correspond to any given subset $S=\{\l_{i_1},\dots,\l_{i_{g-1}}\}$ of $g-1$ arbitrary non-coinciding branch points. The odd half-period associated to the subset $S$ is given by  
\be
\B\pb^S+\qb^S= U(\l_{i_1})+\dots+U(\l_{i_{g-1}})-K.
\la{odd}
\ee
Analogously, we shall be interested in the even half-periods which may be represented as follows,
\be
\B\pb^T+\qb^T = U(\l_{i_1})+\dots+U(\l_{i_{g+1}})-K,
\la{even}\ee
where $T=\{\l_{i_1},\dots,\l_{i_{g+1}}\}$ is an arbitrary subset of $g+1$ branch points.
\begin{theorem}
Let the $2\times 2$ matrix-valued function $\Phi(P)$ be defined on the universal covering $\Lh$ of $\L$ by the following formula,
\be
\Phi(P)=\left(\ba{cc}\phi(P)\;\;\;\;\; \phi(P^*)\\
                  \psi(P)\;\;\;\;\; \psi(P^*)\ea\right),
\la{Psi0}\ee 
where 
\be
\phi(P)=\Th[\pb,\qb](U(P)+U(P_\phi))\Th[\pb^\0,\qb^\0](U(P)-U(P_\phi)),
\la{phi1}
\ee
\be
\psi(P)=\Th[\pb,\qb](U(P)+U(P_\psi))\Th[\pb^\0,\qb^\0](U(P)-U(P_\psi)),
\la{psi1}
\ee
with arbitrary (possibly $\{\l_j\}$-dependent) $P_{\phi,\psi}\in\L$ and arbitrary constant characteristic $[\pb,\qb]$; $*$ is the involution on $\L$ interchanging the sheets. The odd theta characteristic $[\pb^\0,\qb^\0]$ corresponds to an arbitrary subset $S$ of $g-1$ branch points via 
{\rm Eq.~(\ref{odd})}.

Then the function $\Phi(P)$ is holomorphic and invertible outside of the branch points $\l_1,\dots,\l_{2g+2}$ and transforms as follows with respect to the tracing along the  basic cycles of $\L$,
\be
T_{a_j}[\Phi(P)]= \Phi(P)e^{2\pi i (p_j+ p^\0_j)\sigma_3},  
\;\;\;\;\;\;\;
T_{b_j}[\Phi(P)]= \Phi(P)e^{-2\pi i (q_j+  q^\0_j)\sigma_3}
e^{-2\pi i\B_{jj}-4\pi i U(P)},
\la{transPsi}\ee
where by $T_l$ we denote the operator of analytic continuation along the contour $l$. 
Moreover, the function $\Phi$ has the following asymptotic expansion in the neighborhood of point $\l_j$:
\be
\Phi (P)\underset{\lambda\to\lambda_j}=\left\{ F_j+ 
O(\sqrt{\l-\l_j})\right\}\left(\ba{cc}(\l-\l_j)^{1/2+\delta_j}& 0\\
                                 0     & (\l-\l_j)^{\delta_j} \ea \right) 
\left(\ba{cc}1 & 1\\
            1  & -1 \ea \right),
\la{asPsi0}
\ee
with some $\l$-independent matrices $F_j,\,j=1,\dots,2g+2$; $\delta_j=1$ for $\l_j\in S$ and  $\delta_j=0$ for $\l_j\not\in S$.
\la{1}
\end{theorem}
{\it Proof.} 
Let us first check the announced monodromy properties of $\Phi(P)$ around the basic cycles of $\L$. 
{}From the periodicity properties of the theta function given by Eqs.(\ref{pera}),
(\ref{perb}) we deduce the following transformation laws for
$\phi$:
\be
T_{a_j}[\phi(P)]=e^{2\pi i (p_j+p^\0_j)}\phi(P),
\ee
\be
T_{b_j}[\phi(P)]=e^{-2\pi i (q_j+ q^\0_j)}e^{-2\pi i\B_{jj}-4\pi i U(P)} 
\phi(P),
\ee
and the same transformation laws for $\psi$.
Taking into account the action of the involution $*$ on the basic cycles and 
holomorphic differentials, 
\be
a_j^* = -a_j,\;\;\;\;\;\;\;b_j^* = -b_j,\;\;\;\;\;\;\; dU_j(P^*)=-dU_j(P),
\ee
we get the transformation laws for the function
$\phi(P^*)$,
\be
T_{a_j}[\phi(P^*)]=e^{-2\pi i (p_j+p^\0_j)}\phi(P^*),
\ee
\be
T_{b_j}[\phi(P^*)]=
e^{2\pi i (q_j+q^\0_j)}e^{-2\pi i\B_{jj}-4\pi i U(P)} \phi(P^*),
\ee
which coincide with the transformation laws for the function $\psi(P^*)$.
Altogether, this implies relations (\ref{transPsi}) for the
function $\Phi(P)$.

The holomorphy of the function $\Phi$ follows from the holomorphy of the theta function.
Let us show that  $\det\Phi$ does not vanish outside of the branch points $\l_j$. Since the transformations  (\ref{transPsi}) preserve the positions of the zeros of $\det\Phi$, it makes sense to speak about the positions of the zeros of  $\det\Phi$ in the
fundamental polygon $\hat{\L}$. First, notice that 
$\det\Phi(P)$ vanishes at the branch points $\l_j$, where two columns of the matrix $\Phi$ coincide. Moreover, $\det\Phi$ has at the points $\l_j\in S$ zeros of order 3. This can be seen if we rewrite the second theta function in Eq.~(\ref{phi1}) up to a non-vanishing exponential factor as
\ben
\Th(U(P)-U(P_\phi)-\sum_{S} U(\l_j)-K).
\een

Thus we know altogether $3(g-1)+g+3=4g$ zeroes of  $\det\Phi$ taking into account their multiplicities. To check that $\det\Phi$ does not vanish outside of $\l_j$, we integrate the function $\f{\p}{\p\l}\log\det\Phi(P)$ along the boundary of the fundamental polygon $\p\hat{\L}$. {}From the transformation properties (\ref{transPsi}) we deduce
\be
T_{a_j}[\det\Phi(P)]=\det\Phi(P),\;\;\;\;\;\;
T_{b_j}[\det\Phi(P)]=e^{-4\pi i \B_{jj}-8\pi i U_j(P)}\det\Phi(P).
\ee
Now one can check that this integral equals $4g$ in the same way as in the standard calculation of the number of zeros of theta-function of dimension $g$ \cite{Mum}. Therefore $\det\Phi(P)$ does not have any zeros outside of the branch points $\l_j$.
 
The form of the asymptotic expansion (\ref{asPsi0}) is a direct consequence of the holomorphicity of $\phi$ and $\psi$, the structure (\ref{Psi0}) of the function $\Phi$, and the previous discussion of the zeros of $\det\Phi$.  
\vskip0.5cm

Starting from the function $\Phi(P)$ on $\Gamma$ constructed in the Theorem 
\ref{1}, we shall now define a new function $\Psi(Q)$ on the universal covering $X$ of $\C\setminus\{\l_1,\dots,\l_{2g+2}\}$. 
Let us denote by $\Omega\subset\C$ an arbitrary neighborhood of $\infty$ on $\C$ which does not overlap with the points $\l_j$ and the projections of all basic cycles of $\L$ on $\C$. Let us fix some sheet $X_0$ of $X$ choosing the branch cuts between the
points $\l_j$ to lie outside of domain $\Omega$. Let us also fix some sheet $\hat{\L}$ of the universal covering $\Gamma$ of $\L$; then $\hat{\L}$ will contain two non-intersecting copies of $\Omega$. Choose one of them and denote by $\Omega_1$. The domain 
$\Omega_1$ contains the point at infinity, which we call $\infty^1$. Now we are in position to define 
\be
\Psi(\l\in\Omega)=\sqrt{\f{\det \Phi(\infty^1)}{\det \Phi(\l)}}\Phi^{-1}(\infty^1)\Phi(\l)
\la{Psi}\ee
(by $\lambda$ we denote the projection of $Q\in X$ as well as of $P\in \Gamma$ on $\C$). On the rest of $X$ the function $\Psi(Q)$ is defined via the analytic continuation along the contours $l_j$ (Fig.1). 
\begin{theorem}
Let $\pb,\qb\in\C^g$ be an arbitrary set of $2g$ constants such that
$[\pb,\qb]$ is not a half-integer characteristic. Then the function $\Psi(Q\in X)$ defined by {\rm (\ref{Psi}), (\ref{Psi0})} is independent of the choice of the points $P_{\phi,\psi}\in\L$ and the choice of the set $S=\{\l_{i_1},\dots,\l_{i_{g-1}}\}$. Moreover, $\Psi$ is holomorphic outside of the branch points $\l_1,\dots,\l_{2g+2}$, satisfies the normalization conditions $\det\Psi(\l)=1$ and $\Psi(\l=\infty)=I$, and has the anti-diagonal monodromies $M_j$ given by {\rm Eq.~(\ref{eq:CYLINDERS})} 
along the contours $l_j$ 
{\rm (Fig.1)}. The matrix elements of the monodromies {\rm (\ref{eq:CYLINDERS})} are given by the following expressions:
\ben
m_1=i,\;\;\;\;\;\;\;
m_2=i (-1)^{g}\exp\{-2\pi i \sum_{k=1}^g p_k\},
\een
\ben
m_{2j+1}= i (-1)^{g+1}\exp\{2\pi i q_j- 2\pi i \sum_{k=j}^g p_k\},
\een
\be
m_{2j+2}= i(-1)^{g}\exp\{2\pi i q_j-2\pi i \sum_{k=j+1}^g p_k\},
\la{mj}\ee
\la{main}
for $j=1,\dots,g$, where $p_j$ and $q_j$  are components of the 
vectors $\pb$ and $\qb$, respectively.
The asymptotic expansion of $\Psi(Q)$ in the neighborhood of $\l_j$ is of the form {\rm (\ref{eq:PSI_IN_LAMBDA_K})} with some $G_j$ and 
\be
C_j=\f{1}{\sqrt{2i m_j}}\left(\ba{cc} 1 & im \\ -1 & im\ea\right).
\la{Cj}\ee
\la{thPsi}
\end{theorem}
{\it Proof.}
The non-trivial part is to calculate the monodromies $M_j$ of $\Psi(P)$ along the contours $l_j$.

Combining the transformations (\ref{transPsi}) of function $\Phi$ along the basic cycles of $\L$ with the jumps of $\Phi$, 
\ben
\Phi(P)\rightarrow \Phi(P)\sigma_1,
\een
on the branch cuts $[\l_{2j+1},\l_{2j+2}]$, which follow directly from the definition (\ref{Psi0}), we come to the following relations:
\be
\Psi(P) M_{2j+2} M_{2j+1} = \f{T_{l_{2j+1}\circ l_{2j+2}}[\sqrt{\det\Phi(P)}]}
{\sqrt{\det\Phi(P)}}\Psi(P)e^{2\pi i (p_j-p^\0_j)\sigma_3},
\ee
\be
\Psi(P) M_{2j+1} M_{2j}= \f{T_{l_{2j}\circ l_{2j+1}}[\sqrt{\det\Phi(P)}]}
{\sqrt{\det\Phi(P)}}\Psi(P)
e^{2\pi i (q_j-q_{j-1}+q^\0_j-q^\0_{j-1})\sigma_3},
\ee
$j=1,\dots,g$. Furthermore, taking into account that 
\ben
U(\l_1)=0,\;\;\;\;\;\; U(\l_2)=\f12\sum_{k=1}^g\eb_k,
\een
\be
U(\l_{2j+1})= \f12\B\eb_j+ \f12\sum_{k=j}^g\eb_k,
\;\;\;\;\;\;
U(\l_{2j+2})= \f12\B\eb_j+  \f12\sum_{k=j+1}^g\eb_k,\;\;\;\;\;\;
j=1,\dots,g,
\la{Uj}
\ee
we get
\be
p^\0_j=\f12(\delta_{2j+1}+\delta_{2j+2}+1),\;\;\;\;\;\;
q^\0_{j+1}-q^\0_{j}=\f12(\delta_{2j+2}+\delta_{2j+3}+1),
\la{p0q0}\ee
where $\delta_j$ are the same as in Eq.~(\ref{asPsi0}).

The function $\sqrt{\det\Phi(P)}$ transforms in the following way with respect to the tracing along the cycles $l_j$:
\be
T_{l_{2j+1}\circ l_{2j+2}}[\sqrt{\det\Phi(P)}]= 
e^{\pi i (\delta_{2j+1}+\delta_{2j+2}+1)}\sqrt{\det\Phi(P)},  
\la{deta}
\ee
\be
T_{l_{2j}\circ l_{2j+1}}[\sqrt{\det\Phi(P)}]=
e^{\pi i (\delta_{2j+2}+\delta_{2j+3}+1)}\sqrt{\det\Phi(P)}.
\la{detb}
\ee
To prove relations (\ref{deta}), (\ref{detb}) it is enough to notice that in the $\l$-plane the function $\sqrt{\det\Phi(P)}$ has at the point $\l_j$ a zero of degree $3/4$ if $\l_j\in S$ and zero of degree $1/4$ if  $\l_j\not\in S$.

Altogether we get
\ben
M_{2j+2} M_{2j+1}=\exp\{2\pi i p_j\sigma_3\},
\een
\ben
M_{2j+1} M_{2j}=\exp\{2\pi i (q_j-q_{j-1})\sigma_3\},
\een
which imply (\ref{mj}) taking into account that $m_1=i$ and the monodromy around infinity is trivial (\ref{eq:CYLINDERS_CONDITIONS}). 

Now the independence of the function $\Psi$ of the choice of the divisor $S$ and the points  $P_{\phi,\psi}$ follows from the uniqueness of the solution to the Riemann-Hilbert problem with fixed monodromy data.

Existence of the local espansion \Ref{eq:PSI_IN_LAMBDA_K} of the function $\Psi(Q)$ at the points $\l_j$ follows from the related statement (\ref{asPsi0}) for the function $\Phi$  which was proved  in Theorem \ref{1}. The form (\ref{Cj}) of the matrices $C_j$ follows from the relation 
(\ref{eq:A_K_IN_G_K}) between the matrices $M_j$ and $C_j$.
\begin{remark}\rm
The assumption made in Theorem \ref{thPsi} that $[\pb,\qb]$ does not coincide with any half-integer characteristic is nothing but the non-triviality condition, namely, if $[\pb,\qb]$ is a half-integer  characteristic, 
all monodromies $M_j$ become proportional to $\sigma_1$: $M_j=\pm i\sigma_1$; therefore, they can be simultaneously diagonalized by the transformation 
\ben
\Psi\rightarrow \tilde{\Psi}\equiv\Psi \left(\begin{array}{cc}1 & 1\\
                                     1 & -1 \end{array}\right),
\een
The function $\tilde{\Psi}$ has diagonal monodromies $\pm i \sigma_3$, and, therefore, can be chosen to be diagonal itself. Thus, we are in the framework of the scalar Riemann-Hilbert problem: the related matrices $A_j$ are diagonal, and, therefore, $\l_j$-independent, as follows from the 
Schlesinger equations.
\end{remark}

By the special choice   $P_\phi =\infty^2$ and  $P_\psi =\infty^1$ in the formulas of Theorem \ref{1}, we can simplify the previous expression for the function $\Psi$ to get the following
\begin{corollary}
The function $\Psi(\l)$ defined by Eq.~{\rm(\ref{Psi})} may alternatively be represented as follows:
\be
\Psi(\l\in\Omega)=\f{1}{\sqrt{det \Phi^\infty(\l)}}\Phi^\infty(\l),
\la{Psih}\ee
where
\be
\Phi^\infty(P)=\left(\ba{cc}\ph(P)\;\;\;\;\; \ph(P^*)\\
                  \psih(P)\;\;\;\;\; \psih(P^*)\ea\right),
\la{Psih0}
\ee 
\be
\ph(P) = \f{\Th[\pb,\qb](U(P)+U(\infty^2))\Th[\pb^\0,\qb^\0](U(P)-U(\infty^2))}
{\Th[\pb,\qb](0)\Th[\pb^\0,\qb^\0](-2U(\infty^2))},
\la{phinorm}
\ee
\be
\psih(P) = \f{\Th[\pb,\qb](U(P)+U(\infty^1))\Th[\pb^\0,\qb^\0](U(P)-U(\infty^1))}
{\Th[\pb,\qb](0)\Th[\pb^\0,\qb^\0](-2U(\infty^1))}.
\la{psinorm}
\ee
\end{corollary}
{}From the asymptotic expansions of the function $\Phi^\infty(P)$ at the points $\l_j$ we can now construct solutions to the Schlesinger system.
\begin{theorem}
The solution to the Schlesinger system {\rm(\ref{FIRST_SCHLESINGER})}, {\rm(\ref{LAST_SCHLESINGER})} corresponding to the monodromy matrices {\rm(\ref{eq:CYLINDERS})}, {(\rm\ref{mj})} is given by
\be
A_j=\f{1}{4}F_j^{\infty}\sigma_3 (F_j^{\infty})^{-1},\;\;\;
\la{Ajnorm}
\ee
where
\be
(F_j^{\infty})^{11}= \f{\Th[\pb,\qb](U(\l_j)+U(\infty^2))
\Th[\pb^\j,\qb^\j](U(\l_j)-U(\infty^2))}
{\Th[\pb,\qb](0)\Th[\pb^\j,\qb^\j](-2U(\infty^2))},
\la{ajnorm}\ee
\ben
(F_j^{\infty})^{12}= \sum_{k=1}^g \f{\sum_{l=1}^g ({\cal A}^{-1})_{lk} \l_j^{l-1}}{\prod_{l\neq j} (\l_j-\l_l)^{1/2}}
\een
\be
\times\f{\p}{\p z_k}\left\{\f{\Th[\pb,\qb]({\bf z}+U(\infty^2))
\Th[\pb^\j,\qb^\j]({\bf z}-U(\infty^2))}{\Th[\pb,\qb](0)\Th[\pb^\j,\qb^\j](-2U(\infty^2))}\right\}({\bf z}=U(\l_j)),
\la{bjnorm}\ee
and  
$\p/\p z_k$ means the derivative of the theta function {\rm(\ref{theta})} with respect to its $k$th variable; matrix ${\cal A}$ is given by Eq.~{\rm(\ref{AB})}; $\j$ are arbitrary $2g+2$ sets of $g-1$ branch points $\l_j$ satisfying the conditions $\l_j\not\in \j$. The solution {\rm(\ref{Ajnorm})} is independent of the choice of the sets $S_j$ as long as these conditions are fulfilled.

The formulas for the matrix elements $(F_j^{\infty})^{21}$ and $(F_j^{\infty})^{22}$ may be obtained from the formulas for $(F_j^{\infty})^{11}$ and $(F_j^{\infty})^{12}$, respectively, by interchanging $\infty^1$ and $\infty^2$.
\la{Schlesinger}
\end{theorem}
{\it Proof.}
In the neighborhood of the point $\l_j$ we have
\be
\ph_j(P)= (F_j^{\infty})^{11}+\sqrt{\l-\l_j}(F_j^{\infty})^{12}+O(\l-\l_j),
\la{asPh}
\ee
\be
\psih_j(P)= (F_j^{\infty})^{21}+\sqrt{\l-\l_j}(F_j^{\infty})^{22}+O(\l-\l_j),
\la{asPsih}
\ee
with $F_j$ given by Eqs.~(\ref{ajnorm}), (\ref{bjnorm}); the functions $\ph_j(P)$ and $\psih_j(P)$ are defined by Eqs.~(\ref{phi1}), (\ref{psi1}), with $P_\phi =\infty^2$, $P_\psi =\infty^1$, and $[\pb^\0,\qb^\0]$ substituted by $[\pb^\j,\qb^\j]$.

 Therefore, 
\be
\det\Phi_j^{\infty}(P)= \sqrt{\l-\l_j}\{\det F_j^{\infty}+O(\l-\l_j)\},
\ee
and
\ben
[\det \Phi_j^\infty(P)]^{-1/2}\ph_j(P)= [\det F^\infty_j]^{-1}[ (F_j^{\infty})^{11}+\sqrt{\l-\l_j}(F_j^{\infty})^{12}+O(\l-\l_j)],
\een
\ben
[\det \Phi_j^\infty(P)]^{-1/2}\psih_j(P)= [\det F^\infty_j]^{-1}[ (F_j^{\infty})^{21}+\sqrt{\l-\l_j}(F_j^{\infty})^{22}+O(\l-\l_j)].
\een
We conclude that the matrices $G_j$, from the asymptotic expansions (\ref{eq:PSI_IN_LAMBDA_K}) of the function $\Psi(Q)$ at the points $\l_j$, are given by
\be
G_j= (\det F^\infty_j)^{-1} F^\infty_j,
\la{GjFj}\ee
which proves Eq.~(\ref{Ajnorm}).
\begin{remark}\rm
The matrices $F_j^\infty$ from Theorem \ref{Schlesinger} are related to the coefficients $F_j$ of the asymptotic expansions (\ref{asPsi0}) 
of function $\Phi(P)$ at the points $\l_j$ as follows,
\ben
F_j^\infty= \Phi^{-1}(\infty^1) F_j,
\een
Therefore, using Eq.~(\ref{GjFj}), we get the following relation between the matrices $F_j$ from the asymptotic expansions (\ref{asPsi0}) of function $\Phi(P)$ and the matrices $G_j$ from the asymptotic expansions
(\ref{eq:PSI_IN_LAMBDA_K}) of function 
$\Psi(Q)$:
\be
F_k^{-1}F_j\sigma_3 F_j^{-1}F_k=G_k^{-1}G_j\sigma_3 G_j^{-1}G_k, 
\la{FFGG}\ee
for any $j$ and $k$.
\end{remark} 
 
\section{Tau function for the Schlesinger System}
\setcounter{equation}{0}

Here we calculate the $\tau$-function which corresponds to the solution {\rm (\ref{Ajnorm})}, {\rm (\ref{ajnorm})}, {\rm (\ref{bjnorm})} of the Schlesinger system. The remainder is devoted to the proof of 
the following main
\begin{theorem}
The $\tau$-function corresponding to the solution {\rm(\ref{Ajnorm})}, {\rm(\ref{ajnorm})}, {\rm(\ref{bjnorm})} of the Schlesinger system (with arbitrary $\pb,\qb\in\C^g$ such that $[\pb,\qb]$ is not a half-integer characteristic) is given by
\be
\tau= \Theta[\pb,\qb](0)(\det{\cal A})^{-1/2}\prod_{j<k}(\l_j-\l_k)^{-1/8},
\la{tau}\ee
where the $g\times g$ matrix ${\cal A}$ of $a$-periods of  holomorphic 
1-forms on $\L$ is defined by Eq.~{\rm(\ref{AB})}.
\la{thtau}
\end{theorem}
{\it Proof.}
According to the definition of the $\tau$-function (\ref{deftau}), (\ref{DEF_H_J}), let us first calculate $\f{1}{2}\tr (\Psi_\l\Psi^{-1})^2$ for the function $\Psi$ given by Eq.~(\ref{Psi}). We have
\be
\f{1}{2}\tr  (\Psi_\l\Psi^{-1})^2\equiv-\det  (\Psi_\l\Psi^{-1})= 
-\f{\det (\Phi_\l)}{\det \Phi}+
\f{1}{4}\left(\f{(\det\Phi)_\l}{\det\Phi}\right)^2.
\la{mindet}\ee
Together with the function  $\Psi$, the function $\det(\Psi_\l\Psi^{-1})$ is independent of $P_\phi$ and $P_\psi$; moreover, function $\Psi$ does not undergo any modification if we multiply $\psi(P)$ with an arbitrary $\l$-independent factor $C_\psi$. So, we can choose the parameters $P_\phi$, $P_\psi$, and $C_\psi$ at our disposal to simplify the calculation. Our choice will be the following: first we put $C_\psi= \l_\psi-\l_\phi$ ($\l_\phi$ denotes the projection of the point $P_\phi$ in the $\l$-plane)
and then take the limit $P_\psi\rightarrow P_\phi$. We get
\be
\psi(P)=\phi(P)+\f{\partial\phi(P)}{\partial \l_\phi}.
\la{newpsi}\ee
Since the function $\Psi(P)$ is independent of the remaining parameter $P_\phi$, we can calculate
$\det  (\Psi_\l\Psi^{-1})$ assuming $P_\phi= P$.
Intermediate results of this calculation are as follows:
\ben
\f{(\det\Phi)_\l}{\det\Phi}=
2\f{\partial}{\partial\l}\left\{\log\Th[\pb^\0,\qb^\0](-2U(P))\right\},
\een
and
\ben
\f{\det (\Phi)_\l}{\det \Phi}=\f{1}{\Th[\pb,\qb](0)}
\f{\partial^2}{\partial \l\partial \l_\phi}
\left\{\Th[\pb,\qb](U(P)-U(P_\phi))\right\}_{P_\phi=P}
\een
\ben
+\f{1}{\Th[\pb^\0,\qb^\0](-2U(P))}
\f{\partial^2}{\partial \l\partial \l_\phi}
\left\{\Th[\pb^\0,\qb^\0]((-U(P)-U(P_\phi))\right\}_{P_\phi=P};
\een
therefore,
\ben
\f{1}{2}\tr  (\Psi_\l\Psi^{-1})^2(\l)=
-\f{\partial^2}{\partial \l\partial \l_\phi}
\left\{\log\Th[\pb^\0,\qb^\0]((-U(P)-U(P_\phi))\right\}_{P_\phi=P}
\een
\be
-\f{1}{\Th[\pb,\qb](0)}\f{\partial^2}{\partial \l\partial \l_\phi}
\left\{\Th[\pb,\qb](-U(P)+U(P_\phi))\right\}_{P_\phi=P}.
\la{detA}\ee
To find the asymptotic expansion of this expression as $\l\to\l_j$ we shall use the well-known asymptotic expansion which is valid for any odd theta-characteristic $[\pb^\0,\qb^\0]$:
\be
\f{\partial^2}{\partial x(P_1) \partial x(P_2)}\left\{\log\Th[\pb^\0,\qb^\0]
(U(P_1)-U(P_2))\right\}
= \f{1}{(x(P_1)- x(P_2))^2}+ F(P)+ o(1)
\la{ass1}\ee
as $P_1,P_2\rightarrow P$, where $x$ is a local parameter in the neighborhood of $P$. The function $F(P)$ is independent of the choice of the set $S$; it is given by the following expression (\cite{Fay}, p.20),,
\ben
F(P)\equiv \f{1}{6}\{\l,x\}(P)+\f{1}{16}\Big(\f{d}{d x}\log\prod_{k=1}^{g+1}
\f{\l-\l_{i_k}}{\l-\l_{j_k}}\Big)^2 (P)
\een
\be
-\sum_{i,j=1}^g\f{\partial^2}{\partial z_i \partial z_j}\Th[\pb^\1,\qb^\1](0)
\f{dU_i}{d x}(P)\f{dU_j}{d x}(P),
\la{ass2}\ee
where $\{\l,x\}$ denotes the Schwarzian derivative of $\l$ with respect to $x$,
\ben
\f{\l'''}{\l'}-\f{3}{2}\left(\f{\l''}{\l'}\right)^2\;,
\een
and $[\pb^\1,\qb^\1]$ is an even characteristic corresponding to an arbitrary set $\1\equiv \{\l_{i_1},\dots,\l_{i_{g+1}}\}$ of $g+1$ branch points via Eq.~(\ref{even}). The remaining $g+1$ branch points are denoted by $\l_{j_1},\dots,\l_{j_{g+1}}$. Expression (\ref{ass2}) is independent of the choice of the set $\1$.

Applying Eq.~(\ref{ass2}) for $P=\l_j$ we get the following asymptotic expansion, 
\be
\f{1}{2}\tr  (\Psi_\l\Psi^{-1})^2(\l)\underset{\l\to\l_j}=\f{1}{16(\l-\l_j)^2}+ \f{H_j}{\l-\l_j}+ O(1),
\la{ass3}\ee
where
\ben
H_j= \f{1}{8}\sum_{k\neq j}\f{n_j n_k}{\l_j-\l_k}-
\f{1}{4\Th[\pb^\1,\qb^\1](0)}\sum_{l,k=1}^g\f{\partial^2\Th[\pb^\1,\qb^\1]}
{\partial z_l\partial z_k} (0)\f{dU_l}{dx_j}(\l_j)\f{dU_k}{dx_j}(\l_j)
\een
\be
+\f{1}{4\Th[\pb,\qb](0)}\sum_{l,k=1}^g\f{\partial^2\Th[\pb,\qb]}
{\partial z_l\partial z_k} (0)\f{dU_l}{dx_j}(\l_j)\f{dU_k}{dx_j}(\l_j),
\la{Hj}\ee
and $x_j\equiv \sqrt{\l-\l_j}$; $n_k=1$ for $\l_k\in \1$ and  $n_k=-1$
for  $\l_k\not\in \1$.
Now, to integrate Eqs.~(\ref{deftau}), we have to use the heat equations
\be
\f{\partial^2\Th[\pb,\qb](\z|\B)}
{\partial z_l\partial z_k}= 4\pi i\f{\partial\Th[\pb,\qb](\z|\B)}
{\partial \B_{lk}}
\la{heat} 
\ee
valid for theta functions with arbitrary characteristic $[\pb,\qb]$, and the following
\begin{lemma}
The dependence of the matrix of $b$-periods on the branch points is 
described by the following equations,
\be
\f{\partial \B_{kl}}{\partial \l_j}=
\pi i \f{dU_l}{dx_j}(\l_j)\f{dU_k}{dx_j}(\l_j).
\la{Bla}\ee
\end{lemma}
{\it Proof.} 
The dependence of the non-normalized 1-forms $dU_k^0$ (\ref{dUk0}) on $\l_j$ is
\ben
\f{\partial}{\partial \l_j}\{dU_k^0(\l)\} =\f{1}{2(\l-\l_j)}dU_k^0(\l). 
\een
Now, calculation of the integral 
\ben
\oint_{\partial\hat{\L}} U_l^0(\l)\f{\partial}{\partial \l_j}dU_k^0(\l)=
\oint_{\partial\hat{\L}}\f{1}{2(\l-\l_j)} U_l^0(\l)dU_k^0(\l)
\een
by means of the residue theorem gives the following result:
\ben
\pi i \f{dU^0_l}{dx_j}(\l_j)\f{dU^0_k}{dx_j}(\l_j) \equiv
\pi i \left[{\cal A} \f{dU_l}{dx_j}(\l_j)\f{dU_k}{dx_j}(\l_j){\cal A}^t\right]_{kl}.
\een
On the other hand, standard arguments used, for example, in the proof of the Riemann bilinear identities \cite{GurCour}, show that the same integral equals 
\ben
\sum_{m=1}^g {\cal A}_{lm}\f{\partial {\cal B}_{km}}{\partial \l_j}-
\f{\partial {\cal A}_{km}}{\partial \l_j} {\cal B}_{lm}; 
\een
therefore,
\ben
\f{\partial {\cal B}}{\partial \l_j}{\cal A}^t- \f{\partial {\cal A}}{\partial \l_j} {\cal B}^t=
\pi i {\cal A} \f{dU_l}{dx_j}(\l_j)\f{dU_k}{dx_j}(\l_j){\cal A}^t,
\een
which leads to the statement of the lemma  (\ref{Bla}) after taking into account the symmetry of the matrix 
$\B\equiv {\cal A}^{-1} {\cal B}$.
\vskip0.5cm

Now, using Eqs.~(\ref{Hj}), (\ref{heat}), and (\ref{Bla}), we can rewrite the Hamiltonians $H_j$ as follows:
\ben
H_j\equiv\f{\partial }{\partial \l_j}\log\tau= \f{1}{8}\sum_{k\neq j}\f{n_j n_k}{\l_j-\l_k}+\f{\partial }{\partial \l_j}\log
\left\{\f{\Th[\pb,\qb](0)}{\Th[\pb^\1,\qb^\1](0)}\right\}.
\een
Finally, applying the classical Thomae formula \cite{Thomae,Mum}
\ben
\Th^4[\pb^\1,\qb^\1](0)=\pm(\det{\cal A})^2 \prod_{l<k,\;\;l,k=1}^{g+1} (\l_{i_l}-\l_{i_k})
 \prod_{l<k\;\;l,k=1}^{g+1}(\l_{j_l}-\l_{j_k}),
\een
we get the $\tau$-function in the form  (\ref{tau}) up to multiplication by an  arbitrary $\{\l_j\}$-independent constant of integration. The ambiguity in the choice of this constant allows, in particular, to arbitrarily choose the branch cuts in the formula (\ref{tau}).

\section{Elliptic Case and Painlev\'{e} VI Equation}
\setcounter{equation}{0}

In this section we are going to show how the solution of the Painlev\'{e} VI equation in terms of elliptic functions can be derived from the results of the previous sections.
 
Put $g=1$. Then the equation of the curve $\L$ is given by 
\be
w^2=  (\l-\l_1)(\l-\l_2)(\l-\l_3)(\l-\l_4).
\ee
The matrix of $b$-periods, $\B$, turns into the module $\sigma$ and $\Th[\pb^\0,\qb^\0]$ becomes the Jacobi theta-function $\th_1$; to shorten all the formulas we shall denote $\Th[\pb,\qb]$ by $\th_{p,q}$. 

Parameters $m_j$ of the monodromy matrices are, according to (\ref{mj}), given by
\ben
m_1=i,\;\;\;\;\; m_2= -ie^{-2\pi i p},\;\;\;\;\;
m_3=i e^{2\pi i(q-p)},\;\;\;\;\; m_4= -ie^{2\pi i q}.\;\;\;\;\;
\een
The formulas (\ref{phi1}) and (\ref{psi1}) now read as follows
\be
\phi(P)=\th_{p,q}(U(P)+u_\phi)\th_1(U(P)-u_\phi),
\ee
\be
\psi(P)=c_\psi\th_{p,q}(U(P)+u_\psi)\th_1(U(P)-u_\psi),
\ee
where $u_{\phi,\psi}\equiv U(P_{\phi,\psi})\in\C$ are arbitrary parameters, and,  in analogy to the previous section, we introduced an arbitrary multiplier $c_\psi(\{\l_j\})$ which obviously does not influence the function $\Psi(\l)$. 

Again, since the function $\Psi(\l)$ does not depend on $c_\psi$, $u_\phi$ and $u_\psi$, we can freely fix these parameters to simplify our calculations.
First, it is convenient to put $u_\phi=0$ (i.e., $P_\phi=\l_1$), which leads to
\be
\phi(P)=\th_{p,q}(U(P))\th_1(U(P)).
\ee
The most convenient choice for the parameters of the function $\psi$ is the following: we put $c_\psi=u_\psi^{-1}$ and take the limit $u_\psi\rightarrow 0$. Then we get  
\be
\psi(P)= \phi(P)+ \f{\p\phi(P)}{\p u_\phi}(u_\phi=0),\;\;
\la{psin}\ee
and the components of matrices $F_j$ from Eq.~(\ref{asPsi0}) are given by
\ben
F^{11}_j=\th_{p,q}(u_j)\th_1(u_j),
\een
\ben
F^{12}_j= f_j\{\th'_{p,q}(u_j)\th_1(u_j)+\th_{p,q}(u_j)\th_1'(u_j)\}, 
\een
\ben
F^{21}_j=F^{11}_j+\th'_{p,q}(u_j)\th_1(u_j)-\th_{p,q}(u_j)\th_1'(u_j),  
\een
\be
F^{22}_j=F^{12}_j+f_j\{\th''_{p,q}(u_j)\th_1(u_j)-\th_{p,q}(u_j)\th_1''(u_j)\}.\;\;
\la{abcd}
\ee
Here
\be
f_j\equiv
\left\{\prod_{l\neq j} (\l_j-\l_l)^{1/2}\oint_a\f{d\l}{\sqrt{(\l-\l_1)\dots (\l-\l_4)}}\right\}^{-1},
\la{Fj}\ee
and 
\be
u_1=0,\;\;\;\;\; u_2=\f{1}{2},\;\;\;\;\; u_3=\f{1}{2}+\f{\sigma}{2},\;\;\;\;\;
u_4=\f{\sigma}{2}.
\la{Uj1}\ee
In particular, for $j=1$ we have
\be
F^{11}_1=0,\;\;\;\;\;\;\;
F^{21}_1=\th_{p,q}(0)\th_1'(0),\;\;\;\;\;\;\;
F^{12}_1=F^{22}_1 =f_1 F^{21}_1.
\ee
In accordance with Eqs.~(\ref{FFGG}), (\ref{eq:DEF_SOLUTION}), to obtain the solution of the 
sixth Painlev\'{e} equation we have to calculate the $(12)$ elements of the matrices 
\be
\hat{A}_j= \f{1}{4}F_1^{-1}F_j\sigma_3 F_j^{-1} F_1,\;\;\;\;\;\; j=2,3,4
\la{A0j}\ee
(obviously $\hat{A}_1=I$). Substitution of the matrix elements (\ref{abcd})  into Eq.~(\ref{A0j}) 
leads to the following result:
\be
\hat{A}^{12}_j = -f_1
\f{((\log\th_{p,q})'-(\log\th_1)')
({\th''_{p,q}}/{\th_{p,q}}-{\th_1''}/{\th_1})}{(\log\th_{p,q})''-
(\log\th_1)''}(z=u_j),\;\;
\la{A12final}
\ee
where $\th'$ denotes for $\p\th(z|\sigma)/\p z$.
Finally, choosing $\l_1=\infty$, $\l_2=0$, $\l_3=1$, and $\l_4=t$, and making use of the "heat" equation for the theta-function,
\ben
\f{\p\th_{p,q}(z,\sigma)}{\p\sigma}=\f{1}{4\pi i}\f{\p^2\th_{p,q}(z,\sigma)}{\p z^2},\;\;
\een
we get, according to Eq.~(\ref{eq:DEF_SOLUTION}), the following 

\begin{theorem}
The function
\be
y=-\f{t}{1+(1-t) y_1}\;\;,
\la{y1}\ee 
where $t$ is the cross-ratio of the points $\{\l_j\}$ given by Eq.~{\rm(\ref{eq:DEF_T})}, and
\be
y_1=\f{\f{\p}{\p z}\log\f{\p}{\p z}\log\{\th_{p,q}/\th_1\}(\f{1}{2})
\f{\p}{\p \sigma}\log\{\th_{p,q}/\th_1\}(\f{\sigma}{2})}{\f{\p}{\p z}\log\f{\p}{\p z}\log\{\th_{p,q}/\th_1\}(\f{\sigma}{2})
\f{\p}{\p \sigma}\log\{\th_{p,q}/\th_1\}(\f{1}{2})}\;\;.
\la{solution}\ee
where $p,q\in\C$ are arbitrary constants such that $[p,q]\neq [1/2,0]$ and  
$[p,q]\neq [0,1/2]$,
solves the sixth Painlev\'e equation {\rm (\ref{eq:P6})} , with coefficients 
{\rm(\ref{eq:OKAMOTO_COEFFICIENTS})}. Here the module  $\sigma$ of elliptic curve $\L$ is chosen
such that $t=\theta_4^4(0|\sigma)/\theta_2^4(0|\sigma)$.
\end{theorem}

Expression (\ref{solution}) is a combination of derivatives of the function
$\log\f{\th_{p,q}}{\th_1}$ with respect to both arguments of the theta functions.

One more representation for solution (\ref{y1}) of sixth Painlev\'e equation may be obtained by using the following relation between $y(t)$ and the $\tau$-function, $\tau(t)$, valid for $t_j=\f12$:
\be
  y(t)=t-t(t-1)\left[D\left(\frac{\frac d{dt}D(\mathbf{\tau})}
  {\frac d{dt}D\left(\sqrt[8]{t(t-1)}\mathbf{\tau}\right)}\right)
+\frac{t(t-1)}{D^2\left(\sqrt[8]{t(t-1)}\mathbf{\tau}\right)}\right]^{-1}
\la{solut}
\ee
where operator $D$ acts on functions $f(t)$ as follows:
$D(f)\equiv\f{d}{dt}\log f$.
The $\tau$-function for the $g=1$ case can be obtained from the general formula
(\ref{tau}) simply by assuming that $\l_1,\dots,\l_4$ coincide with $0,1,t$, and $\infty$, respectively. Then up to an arbitrary constant we get
\ben
\mathbf{\tau}(t)=\f{\theta_{p,q}(0\vert\sigma)}{\sqrt[8]{t(t-1)}}
\left[\int_0^1\f{d\l}{\sqrt{\l(\l-1)(\l-t)}}\right]^{-\f12}.
\een
\begin{remark}\rm 
It seems that it is not easy to check directly (by applying appropriate identities for the theta functions) the coincidence of the different forms of the same solution
(\ref{solution}), (\ref{solut}). 
It is also not easy to check directly coincidence of our formulas to
other forms of this solution 
given by Okamoto (\ref{eq:SOLUTION_OKAMOTO})  and  Hitchin
(\ref{eq:HITCHIN_THETA}). However, we can explicitly see the relationship of our construction to the construction by Hitchin on the level of the functions $\phi$ and $\psi$ from Theorem \ref{1}, namely, the choice of the rows of the function $\Phi$ made in \cite{H} corresponds to the choice
$u_\phi\equiv-\f{1}{2}(p\sigma+q)+\f{\sigma+1}{4}$. The variable $c$ from \cite{H} is given by $-u_\phi w_1$, where $w_1$ is the first full elliptic integral on $\L$. The parameter $u_\psi$ is fixed in \cite{H} to coincide with one of the zeros of the 
Weierstrass $\wp$-function, $\wp[w_1(U(P)+ u_\phi)]$, with the periods $w_1$ and $w_2=w_1\sigma$.
Constants $c_1$ and $c_2$ from \cite{H} are related to our $p$ and $q$ as follows: $c_1=p+\f{1}{2}$, $c_2=q+\f{1}{2}$.
\end{remark}

\begin{remark}\rm
Here we discussed only generic two-parametric family of elliptic
solutions of Painlev\'e 6 equation with coefficients 
(\ref{eq:OKAMOTO_COEFFICIENTS}), which corresponds to monodromy matrices
(\ref{eq:SET_GENERAL}). Additional one-parametric family of solutions corresponding to monodromy matrices (\ref{eq:SET_SPECIAL}) was constructed in \cite{H}.
\end{remark}

\begin{appendix}

\section{Elliptic Solutions of the Sixth Painlev\'e Equation}
\setcounter{equation}{0}
In his studies of the Painlev\'e equations K.~Okamoto has shown \cite{O}
that
the function $y=y(t)$, the general solution of the sixth Painlev\'e
equation, (\ref{eq:P6}), can be
explicitly written in terms of the elliptic functions provided the set of
the parameters satisfies one of the following conditions:

\begin{equation}
  \label{eq:ELLIPTIC_CONDITIONS_0}
t_i\in\Bbb Z,\;\;\;\;\;\;t_1+\dots+t_4\in 2\Bbb Z,
\end{equation}
or
\begin{equation}
  \label{eq::ELLIPTIC_CONDITIONS_1}
t_i+\f12\in \Z.\;\;\; 
\end{equation}
The major ingredients of the Okamoto's construction are:\\
1) The so-called Picard solution, 
\begin{equation}
  \label{eq:PAINLEVE}
  y_0(t)=\tilde\wp(c_1\omega_1(t)+c_2\omega_2(t)),
\end{equation}
of Eq.~(\ref{eq:P6}) with the coefficients:
\begin{equation}
  \label{PARAMETERS_ZERO}
\alpha=0,\quad\beta=0,\quad\gamma=0,\quad\delta=\frac 12.
\end{equation}
In Eq.~(\ref{eq:PAINLEVE}) $\tilde\wp(\cdot)$ is the elliptic function
satisfying the equation, 
$\tilde\wp^{'\,2}=4\tilde\wp(\tilde\wp-1)(\tilde\wp-t)$,
with the primitive periods $2\omega_1(t)$ and 
$2\omega_2(t)$;
$c_1,\,c_2\in\Bbb C$ are the constants of integration, so that the
function $y(t)$ is the general solution.
\newline 
2) The subgroup of transformations of solutions of Eq.~(\ref{eq:P6}) which
acts on the space of coefficients $\{t_j\}$ 
as: a)reflections: for any $j=1,\dots,4$ there is a transformation which 
transforms $t_j\to-t_j$ and leaves all $t_k$ for $k\neq j$ unchanged;
b) permutations of the set $\{t_j\}$;
c) the shifts:  
$t_j\mapsto t_j+n_j$,
where $\sum_{j=1}^4 n_j=0({\rm mod}\;2)$. 
\newline
3) More nontrivial transformation,
\begin{eqnarray}
  \label{eq:OKA_TRNS}
{\bf O}:&(t_1,\,t_2,\,t_3,\,t_4)\leftrightarrow 
\left(\frac{t_1+t_2-t_3-t_4}2,\,\frac{t_1+t_2+t_3+t_4}2,\,
\frac{-t_1+t_2+t_3-t_4}2,\,\frac{-t_1+t_2-t_3+t_4}2\right).&  
\end{eqnarray}
It is important to mention that all the transformations described above,
as well as their inversions,
are given by explicit formulas, so that ``new'' solutions can be explicitly
written in terms of the ``old'' ones as rational functions of the ``old'' 
solution and its derivative (see \cite{O}). In particular, the solution of
Eq.~(\ref{eq:P6}) with the coefficients (\ref{eq:OKAMOTO_COEFFICIENTS})
obtained via the Okamoto's transformations reads
\begin{equation}
  \label{eq:SOLUTION_OKAMOTO}
  y(t)=y_0+\frac{y_0^2(y_0-1)(y_0-t)}{t(t-1)y_0^{\prime}-y_0(y_0-1)},
\end{equation}
where $y_0=y_0(t)$ is given by Eq.~(\ref{eq:PAINLEVE}).

N.~Hitchin, in the work \cite{H} devoted to the study of $SU(2)$-invariant 
anti-self-dual Einstein metrics, rediscovered the case
(\ref{eq:OKAMOTO_COEFFICIENTS})  
of integrability of Eq.~(\ref{eq:P6}) in elliptic functions. He got the
following representation for the solution (\ref{eq:SOLUTION_OKAMOTO}) in the parametric form, 
\begin{eqnarray}
  \label{eq:HITCHIN_THETA}
&\!\!\!y_1(\sigma)=\!\frac{\theta^{'''}_1\!(0)}
{3\pi^2\theta^4_4(0)\theta^{'}_1\!(0)}+
\frac 13\!\left(1+\frac{\theta^4_3(0)}{\theta^4_4(0)}\right)+
\frac{\theta^{'''}_1\!(\nu)\theta_1(\nu)-
2\theta^{''}_1\!(\nu)\theta^{'}_1\!(\nu)+
2\pi ic_1(\theta^{''}_1\!(\nu)\theta_1(\nu)-{\theta^{'}_1}^2\!(\nu))}
{2\pi^2\theta^4_4(0)\theta_1(\nu)(\theta^{'}_1\!(\nu)+
\pi ic_1\theta_1(\nu))},&\phantom{22}\\
\phantom{2)}&\!t(\sigma)=\frac{\theta^4_3(0)}{\theta^4_4(0)},\quad\quad
\nu=c_1\sigma+c_2,\phantom{hhhhhhhhhhhhhhhhhhhhhhhhhhhhhhhhhh}\nonumber
\end{eqnarray}
where $\theta_k(\cdot)=\theta(\cdot|\sigma),\;k=1,\dots,4,$  are the
Jacobi theta functions \cite{WW}.\\

Yu.~I.~Manin \cite{M} noticed that the well-known uniformization of the
Eq.~(\ref{eq:P6}) in terms of the Weierstrass $\wp$-function can be
further converted into the beautiful form:
\begin{eqnarray}
  \label{MANIN_FORM}
y(\sigma)\!\!\!&=&\!\!\!\frac{\wp(z(\sigma),\sigma)-e_1(\sigma)}{e_2(\sigma)-e_1(\sigma)
},
\phantom{hhhhhhhhhhhhhhhi}
t(\sigma)=\frac{e_3(\sigma)-e_1(\sigma)}{e_2(\sigma)-e_1(\sigma)},\nonumber\\
e_j(\sigma)\!\!\!&=&\!\!\!\wp(\frac 12T_j,\sigma),\;\phantom{hhhhhhhhhhhhhh}
(T_1,T_2,T_3,T_4)\equiv(0,1,\sigma,1+\sigma),\nonumber\\
\frac{d^2z}{d\sigma^2}\!\!\!&=&\!\!\!\frac 1{(2\pi i)^2}
\sum\limits_{j=1}^4\alpha_j\wp^{\prime}(z+\frac{T_j}2,\sigma),\quad
(\alpha_1,\alpha_2,\alpha_3,\alpha_4)
\equiv(\alpha,-\beta,\gamma,\frac 12-\delta),
\end{eqnarray}
where $\wp(\cdot,\sigma)$ is the Weierstrass elliptic function with the
primitive periods $2$ and $2\sigma$;
$\wp^{'}(\cdot,\sigma)$ denotes partial derivative of $\wp$-function with respect to its first argument.
By applying to 
Eq.~(\ref{MANIN_FORM}) the Landin transform for the Weierstrass elliptic 
functions Manin found a new transformation for solutions of
Eq.~(\ref{eq:P6}). In terms of the Manin variables, $z$ and $\sigma$ this 
transformation reads: 
if $z(\sigma)$ is any solution of Eq.~(\ref{MANIN_FORM}) with the
coefficients
$\alpha_1=\alpha_3,\;\alpha_2=\alpha_4$, then $z(2\sigma)$ is the solution of
Eq.~(\ref{MANIN_FORM}) for $\alpha_1^{new}=4\alpha_1,\;
\alpha_2^{new}=4\alpha_2,\;\alpha_3^{new}=\alpha_4^{new}=0$. 
The converse statement is, of course, also true. Schematically,
for the constants, $t_j$ (\ref{DEF_T_J}), we can write: 
\begin{equation}
  \label{eq:MANIN_TRANSFORM}
  {\bf M}:\quad(t_1,\,t_2,\,t_3=t_1-1,\,t_4=t_2)
\leftrightarrow(2t_1-1,\,2t_2,\,0,0).
\end{equation}
In the case (\ref{PARAMETERS_ZERO}) the Manin form of the sixth Painlev\'e 
equation (\ref{MANIN_FORM}) immediately reproduces the Picard solution
(\ref{eq:PAINLEVE}).
In terms of the parameters $t_j$ Eqs.~(\ref{PARAMETERS_ZERO}) read,
$t_1=1,\,t_2=0,\,t_3=0$, and $\,t_4=0$. After the permutation we get
the set $t_1=0,\,t_2=1,\,t_3=0$, and $\,t_4=0$, therefore,
by setting the formal monodromies $t_1=\frac 12,\,t_2=-\frac 12$ 
in the r.-h. s. of (\ref{eq:MANIN_TRANSFORM}) and
choosing the left arrow in Manin transformation {\bf M},
one finds the second basic case of the integrability 
(\ref{eq:OKAMOTO_COEFFICIENTS}). The corresponding explicit formula
can be written as the composition of the transformation
corresponding to the permutation \cite{O} and {\bf M}.

\end{appendix}

\end{document}